\documentclass[prb,twocolumn,amsmath,amssymb,showpacs]{revtex4}
\usepackage{amsfonts}
\usepackage{amsmath}
\usepackage{dcolumn}
\usepackage[dvips]{graphicx,color}
\begin{document}
\title{Critical Temperatures of a Two-Band Model for Diluted Magnetic Semiconductors}
\author{F. Popescu$^1$, Y. Yildirim$^1$, G. Alvarez$^2$ A. Moreo$^{3,4}$,
and E. Dagotto$^{3,4}$}
%\footnote{florentin_p@hotmail.com}

\affiliation{$^1$Physics Department, Florida State University,
Tallahassee, FL 32306,}

\affiliation{$^2$ Computer Science and Mathematics Division, Oak
Ridge National Laboratory, Oak Ridge, TN 37831-6032,}

\affiliation{$^3$ Department of Physics, The University of
Tennessee, Knoxville, TN 37996,}

 \affiliation{$^4$ Condensed Matter Sciences Division, Oak Ridge
National Laboratory, Oak Ridge, TN 37831-6032}

\begin{abstract}
Using Dynamical Mean Field Theory (DMFT) and Monte Carlo (MC)
simulations, we study the ferromagnetic transition temperature
($T_{c}$) of a two-band model for Diluted Magnetic
Semiconductors (DMS), varying coupling constants, hopping
parameters, and carrier densities. We found that $T_{c}$ is
optimized at all fillings $p$ when both impurity bands (IB) fully overlap in
the same energy range, namely when the exchange couplings $J$ and bandwidths
are identical. The optimal $T_{c}$ is found to be about twice larger
than the maximum value obtained in the one-band model, showing the
importance of multiband descriptions of DMS at intermediate $J$'s.
\end{abstract}
\maketitle

\section{Introduction}
DMS are semiconductors where a fraction $x(\sim$0.01-0.1) of
nonmagnetic elements is replaced by magnetic $\mathrm{Mn}$ ions. The
class of III-V DMS has recently attracted considerable attention
after the experimental observation of high  $T_{c}$'s, due to
significant improvements in molecular beam epitaxy techniques.
\cite{OHN96} These compounds can play an important role in
spintronic devices, and they represent a challenge to theory due to
the combined presence of correlations and disorder.

Using a one-band model for DMS, (i) the weak-coupling quadratic
dependence of $T_{c}$ with $J$ is captured correctly by DMFT
\cite{CHA01}, which also provides bulk limit results, while (ii) the
MC techniques on finite clusters properly handle the random Mn
distribution and unveiled the reduction of $T_c$ at large $J$ due to
carrier localization.\cite{ALV03} For $J$ comparable to the hopping
$t$, both techniques reached similar conclusions, showing that they
can complement towards a comprehensive understanding of DMS.
However, much theoretical work remains to be done, particularly the
consideration of the several bands active in real DMS materials
beyond mean field approximations.\cite{SIN02}

In this paper, we study a two-band model for DMS at realistic low
dopings $x$$\ll$$1$, using a combination of $nonperturbative$
techniques that gives unique characteristics to our work. To
properly analyze the intermediate and strong coupling $J$ regime, MC
simulations that correctly handle the random Mn distribution are
crucial. Here, MC results for DMS models including more than one
band are presented for the first time.\cite{MC-before} On the DMFT
side, pioneering calculations for two-band double-exchange
Hamiltonians already exist.\cite{MOR05} However, those calculations
focused on special cases, while our current DMFT effort is more
general, with two $s$=1/2 bands and arbitrary couplings, hoppings,
and carrier densities. One of our main results is that $T_{c}$ can
be substantially raised by considering multiband systems since at
intermediate couplings the maximal $T_{c}$ at carrier filling
$p$$\cong$$x$ is approximately twice larger than the highest $T_c$
obtained in the single-band model at filling $p$$\cong$$x/2$ (note
that the alternative notation $p_h$=$p/x$, where $p_{h}$ is the hole
density as a fraction of $p$, will be used in some portions of the
paper). The  qualitative reason is that the IB cooperate to raise
$T_{c}$ for values of the chemical potential $\mu$ where those bands
partially or fully overlap.\cite{clarification}

The paper is organized as follows: Section II describes the model; the DMFT
results are presented in Section III, while Section IV is devoted to the
Monte Carlo study. Conclusions and final remarks are in Section V.

\section{Model}
It is known that in $\mathrm{Mn}$-doped $\mathrm{GaAs}$, the
$\mathrm{Mn}$ ions substitute for $\mathrm{Ga}$ cations and
contribute itinerant holes to the valence band. The $\mathrm{Mn}$
ions have a half-filled $d$-shell which acts as a $S=5/2$ local
moment. Due to a strong spin-orbit (SO) interaction, the angular
momentum $\mathbf{L}$  of the $p$-like valence bands mixes with the
hole spin degree of freedom $\mathbf{s}$  and produces low- and
high-energy bands with angular momentum $j=1/2$ and $3/2$,
respectively. A robust SO split between these bands causes the holes
to populate the $j=3/2$ state, which itself is split by the crystal
field into a $m_{j}=\pm3/2$ band with heavy holes and a
$m_{j}=\pm1/2$ band with light holes. This is the reason why we
choose to study two bands since this is the relevant number of
orbitals in most III-V DMS. Since we do not work in a $(j,m_j)$
basis our Hamiltonian does not capture the orbital mixing in the
Hund term.\cite{zarand} However, we roughly consider the diagonal SO
effects in the magnetic interactions by allowing different values of
$J$ in the two orbitals considered. The simple two-band model for
DMS used here is given by the Hamiltonian
\begin{eqnarray}\label{ham}
{\mathcal{H}}\!=-\!\sum_{l,ij,\alpha} t_l
(c^{+}_{l,i,\alpha}c_{l,j,\alpha}+\textrm{H.c.})
\!-\!\!\sum_{l,I} J_l
\mathbf{S}_{I}\cdot\mathbf{s}_{l,I},
\end{eqnarray}
where $l$=1,2 is the band index (not to be confused with angular
momentum), $i,j$ label sites (nearest neighbors for the hopping
term), $c_{l,i,\alpha}$ creates a hole at site $i$ in the band $l$,
$\mathbf{s}_{l,i}=\sum_{\alpha,\beta}
c^{+}_{l,i,\alpha}\mathbf{\sigma}_{\alpha\beta}\,c_{l,i,\beta}$ is
the spin- operator of the mobile hole ($\hat{\mathbf{\sigma}}$ =
Pauli vector), $\alpha$ and $\beta$ are spin indices, $J_{l}$ is the
coupling between the core spin and the electrons of band $l$, and
$\mathbf{S}_{I}$ is the spin of the localized $\mathrm{Mn}$ ion at
randomly selected sites $I$, assumed classical in the MC
simulations. $t_{l}$ is the hopping term in band $l$. The inter-band
hopping $t_{12}$ ($=t_{21}$) is zero at the nearest neighbor level
in cubic lattices \cite{inprogress}. Even if $t_{12}$ is explicitly
added, the conclusions are similar as reported here \cite{details}.
While real DMS materials have zincblende (ZB) structures, in this
first nonperturbative study of a multiband DMS model the simplicity
of a cubic lattice allows us to focus on the dominant qualitative
tendencies, a first step toward future quantitative studies with
realistic ZB lattices.
%(NOTE: IT APPEARS WHEN DIAGONAL HOPPINGS ARE ADDED)
The model will be studied using DMFT and MC techniques.
\begin{figure}
{\scalebox{0.32}{\includegraphics[angle=0]{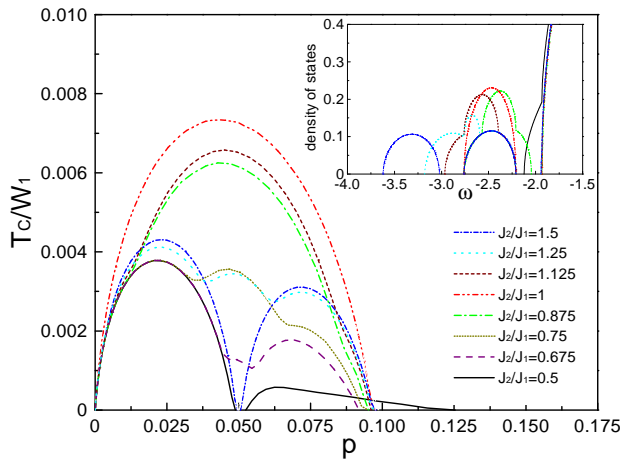}}}
\caption{(Color online) $T_{c}$ versus the carrier concentration
$p$, at various $J_{2}/J_{1}$, obtained with the DMFT technique.
Here, $x=0.05$, $W_{1}/W_{2}=1$, and $J_{1}/W_{1}=0.5$. The inset
shows the corresponding DOS at $T=0$.}\label{Fig1}
\end{figure}
%---------DMFT Results--------------
\section{DMFT Results}
\subsection{Formalism and tests}
Within DMFT the self-energy is assumed to be local,
$\Sigma(\mathbf{p},i\omega_{n})\rightarrow\Sigma(i\omega_{n})$,
assumption valid exactly only in infinite dimensions. The
information about the hopping of carriers on and off lattice sites,
which are magnetic (with probability $x$) or nonmagnetic (with
probability $1$-$x$), is in the bare Green's function (GF)
${\mathcal{G}}_{0}(i\omega_{n})$ \cite{MOR05}. The full GF
${\mathcal{G}}(i\omega_{n})$ can be solved by integration with the
result:
$\langle{\mathcal{G}}(i\omega_{n})\rangle=x\langle[{\mathcal{G}^{-1}_{0}}
(i\omega_{n})+J\mathbf{S}{\mathbf{m}}\hat{\sigma}]^{-1}
\rangle+(1-x)\langle{\mathcal{G}}_{0}(i\omega_{n})\rangle$, where
$\omega_{n}$=$(2n+1)\pi T$ are Matsubara frequencies. The average
$\langle X(\mathbf{m})\rangle$=$\int
d\Omega_{m}X(\mathbf{m})P(\mathbf{m})$ is over the local moment
$\mathbf{m}$ orientations, with probability $P(\mathbf{m})$. The
equation above was derived for each band, and  the equations set for
the GF were solved on a Bethe lattice with a bare semicircular
noninteracting density of states
$\mathrm{DOS}$=$8(W^{2}_{l}/4-\varepsilon^{2})^{1/2}/\pi W^{2}_{l}$.
In this case,
$\langle{{\mathcal{G}}}^{-1}_{0,l}(i\omega_{n})\rangle=z_{n}-(W^{2}_{l}/16)\,\langle\mathcal{G}_{l}(
i\omega_{n})\rangle$, where $z_{n}$=$i\omega_{n}+\mu$, and
$W_{l}$=$4t_{l}$. $T_c$ is obtained by linearizing the bare inverse
GF with respect to the local order parameter $M=\langle m_{z}
\rangle_{\mathbf{m}}$. To first order in $M$, the implicit equation
for $T_{c}$ is:
\begin{widetext}
\begin{equation}\label{TC}
-\sum_{l=1}^{2}\sum_{n=0}^{\infty}\frac{4xJ^{2}_{l}W_{l}^{2}B^2_{l}(i\omega_{n})}{48[B^{2}_{l}(i\omega_{n})-J_{l}^{2}]^{3}
+(48J_{l}^{2}-3W^{2}_{l})[B^{2}_{l}(i\omega_{n})-J_{l}^{2}]^{2}-5xJ_{l}^{2}W_{l}^{2}[B^{2}_{l}(i\omega_{n})-J_{l}^{2}]-2xJ_{l}^{4}W_{l}^{2}}=1,
\end{equation}
\end{widetext}
where $B_{l}(i\omega_{n})$ satisfies a fourth-order-degree equation:
$B^{4}_{l}-z_{n}B^{3}_{l}-(J^{2}_{l}-W_{l}^{2}/16)B^{2}_{l}+z_{n}J^{2}_{l}B_{l}-(1-x)J^{2}W^{2}_{l}/4=0$.\cite{details}
Our formula was tested in different cases: (1) for $J_{2}$=$0$, we
reproduce the one-band results of Ref.\onlinecite{CHA01}, and (2) at
$x$=1, $J_2$=0, and $J_1 \rightarrow \infty$ we reproduced the
results of Ref. \onlinecite{FIS03}. $T_{c}$, contained in the
Matsubara frequencies, is obtained from Eq.(\ref{TC}) numerically.
The equations for $B_{l}$, rewritten in real frequency space via
$i\omega_{n}\rightarrow\omega$, give the interacting $\mathrm{DOS}$
for each band at zero temperature when $\mu=0$.\cite{DOS} The
solutions of these equations depend crucially on the ratio
$J_{l}/W_{l}$. The critical value for the formation of well-defined
IB, corresponding to carrier spins locally parallel to $\mathrm{Mn}$
spins, is $J_{l}/W_{l}\sim 0.33$. Below, the domain
$J_{l}/W_{l}<0.33$ is referred to as  `weak coupling', while
$0.33\lesssim J_{l}/W_{l}\lesssim0.5$ is the `intermediate
coupling'. The most interesting physics is observed at the boundary
between these two regimes, i.e. when the IB are not completely
separated from the valence bands.
\subsection{DMFT critical temperatures varying exchange couplings}
\begin{figure}
{\scalebox{0.32}{\includegraphics{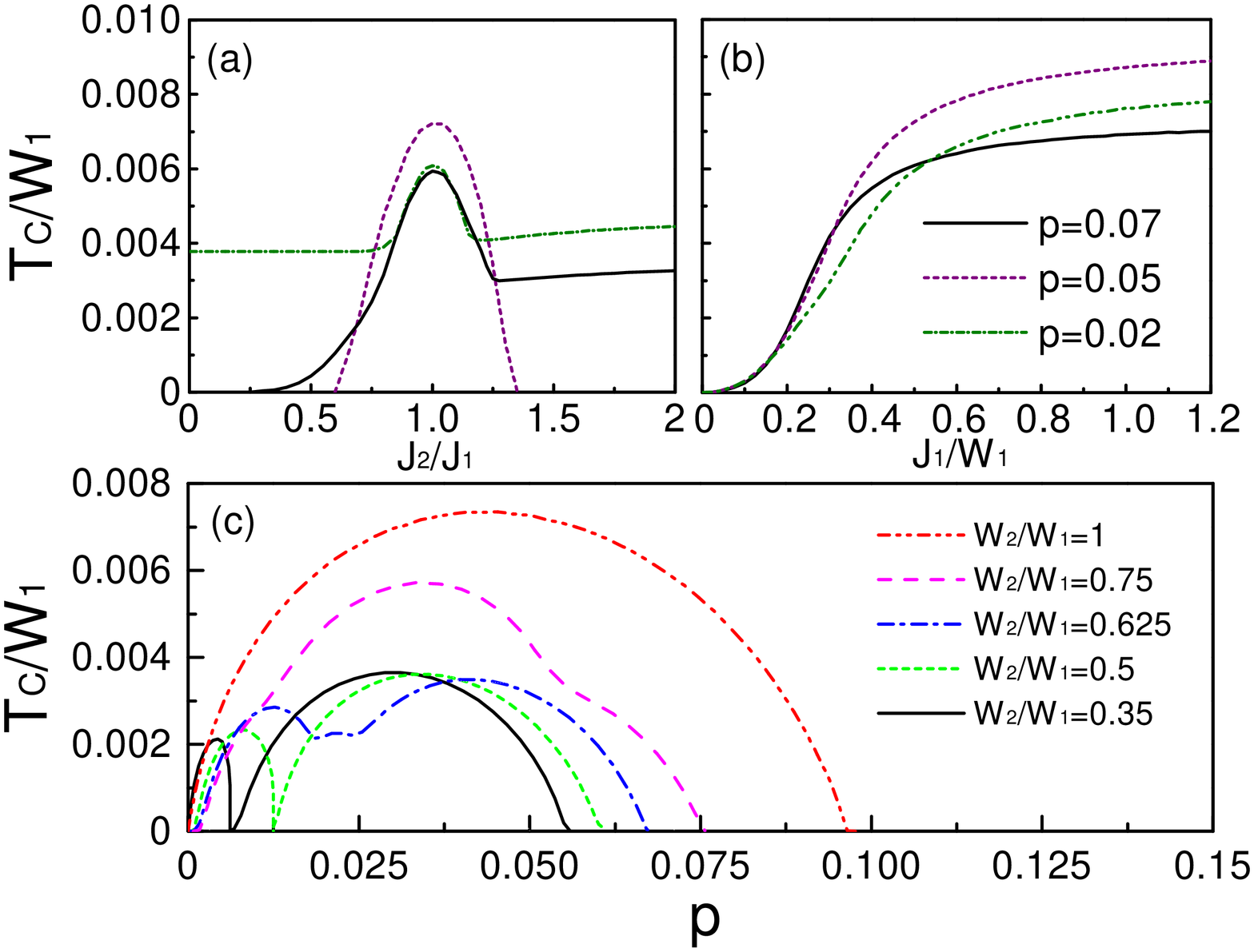}}} \caption{(Color
online) Results obtained with the DMFT approximation: (a) $T_{c}$
vs. $J_{2}/J_{1}$, at $W_{1}/W_{2}=1$, and $J_{1}/W_{1}=0.5$, for
the values of $p$ indicated in (b). At $p=0.02$ and $J_{2}/J_{1}=0$,
a finite $T_c/W_1\sim0.0037$ is caused by the $l$=1 band. At
$p=0.05$, $T_{c}$ is not zero for $J_{2}/J_{1}\in(0.6,1.35)$, and it
increases significantly due to band overlap. The case $p=0.07$
($>x$) corresponds to the first IB completely filled. (b) $T_{c}$
vs. $J_{1}/W_{1}$, at $W_{1}/W_{2}=1$ and $J_{1}/W_{1}=J_{2}/W_{2}$,
for the $p$'s indicated. (c) $T_{c}$  vs. $p$ at different ratios
$W_{2}/W_{1}$, fixing $J_{1}=J_{2}=2$. In all the frames
$x=0.05$.}\label{Fig2}
\end{figure}

In Fig.~\ref{Fig1}, we show $T_{c}$ vs. $p$, for different ratios
$J_{2}/J_{1}$ and at fixed $W_2/W_1$=1 and $J_{1}/W_{1}=0.5$,
situation corresponding to the existence of a well-defined $l$=$1$
IB (although $p$$\leq$$x$ in real DMS, in this paper the case
$p$$>$$x$ will also be studied for completeness, as done in
Ref.\onlinecite{CHA01}). The inset shows the total interacting DOS
evolution. The IB overlap if $|J_{2}/W_{2}-J_{1}/W_{1}|<0.5$. If the
IB do not overlap, then each one determines $T_c$ separately,
causing the double-peak structure observed for some $J_2/J_1$
ratios. The band with the largest $J_{l}/W_{l}$ is filled first, for
smaller $\mu$'s. At all $p$'s, we found that $T_{c}$ is maximum when
$J_{2}/J_{1}=1$, namely when the IB fully overlap. The dependence of
$T_{c}$ on $J_{2}/J_{1}$ at fixed $p$ is in Fig.~\ref{Fig2}(a). The
peak value is achieved when the bands fully overlap (i.e. at
$J_{2}/J_{1}$=1). Once the bands decouple, the value for $T_{c}$
matches one-band model results.
\subsection{DMFT critical temperatures varying bandwidths}
Let us consider now how changes in bandwidths influence $T_{c}$. In
Fig.\ \ref{Fig2}(c), we show $T_{c}$ vs. $p$ parametric with
$W_{2}/W_{1}$, at fixed $J_{1}/W_{1}=0.5$ (intermediate coupling),
and with $J_{2}/J_{1}=1$.\cite{comment} At small $W_{2}/W_{1}$ the
second IB shall be located in a region of $\omega$ smaller (i.e.
farther from the valence bands), than the energy interval occupied
by the $l=1$ IB. Hence, the $l$=2 IB will be the first to be filled.
Decreasing $J_{2}/W_{2}$, the second band moves to the right on the
$\omega$ axis, towards the location of the first band. While the
bands are still separated, each gives its own contribution to
$T_{c}$. The curves with $W_{2}/W_{1}=0.35$ and $0.5$ correspond to
decoupled IB, while those with $W_{2}/W_{1}=0.625$ and $0.75$
correspond to partially overlapping bands. Again, $T_{c}$ is
maximized at all fillings when the bands fully overlap
($W_{2}/W_{1}$=$1$), in good agreement with Ref.\onlinecite{MOR05}.
Although $W_{2}/W_{1}$=$1$ is not realistic in DMS, Mn doping
materials with a relatively small heavy-light mass ratio  will favor
a higher $T_c$. \vskip 1cm
\begin{figure}
{\scalebox{0.32}{\includegraphics{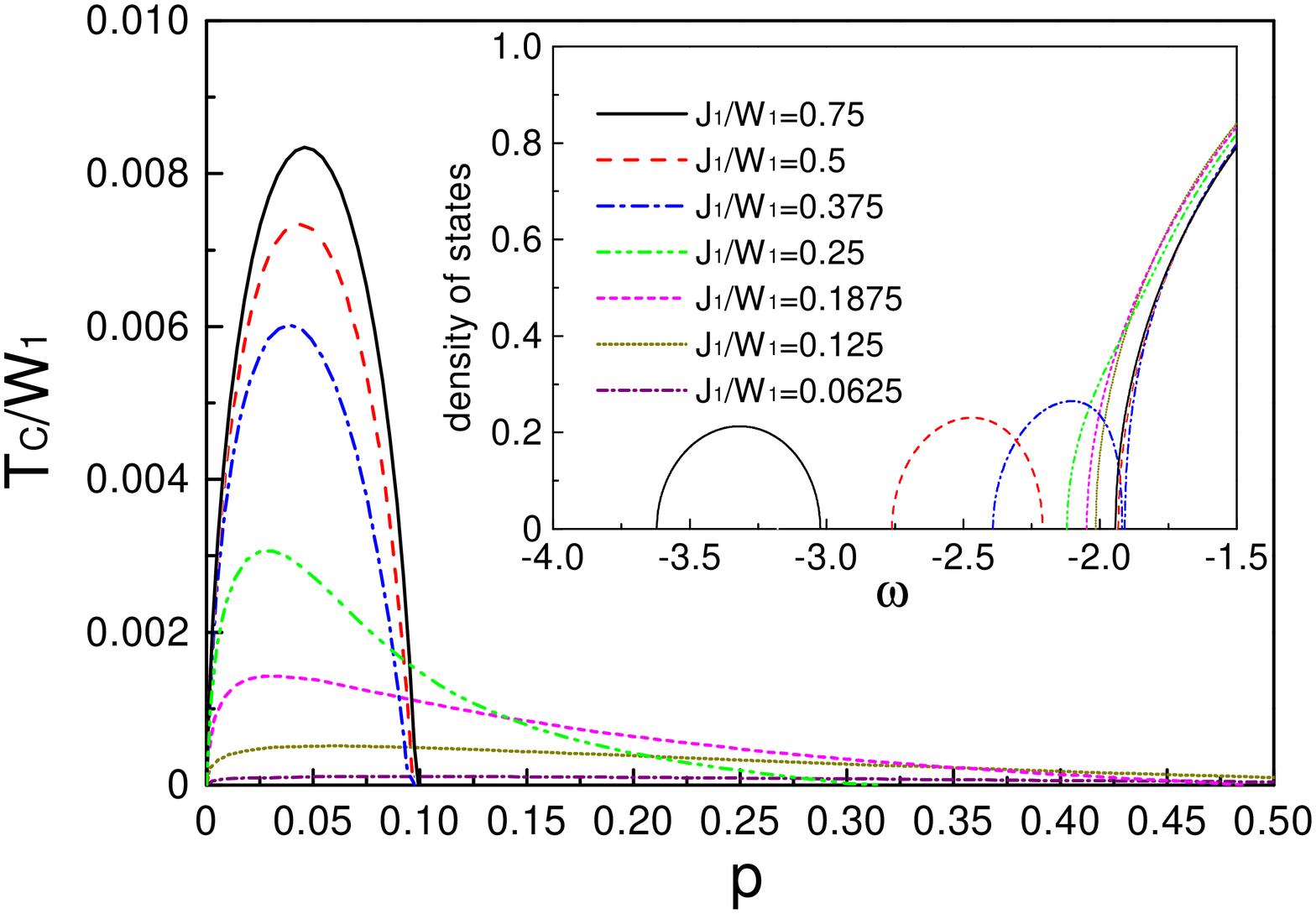}}} \caption{(Color
online) $T_{c}$ versus $p$ at different ratios $J_{1}/W_{1}$,
obtained using DMFT. The parameters $x$, $W_{1}$, $W_{2}$, and
$J_{1}$ are as in Fig.~1. The inset contains the DOS at $T=0$.}
\label{Fig3}
\end{figure}
\subsection{Critical temperatures varying the exchange to bandwidth ratio}
Once established that $T_{c}$ is maximal for all $p$ when
$J_{2}/J_{1}$=$1$ and $W_{2}/W_{1}$=$1$, let us analyze $T_{c}$ vs.
$p$ when $J_{1}/W_{1}$ varies. The results for $T_{c}$, and total
interacting DOS, are in Fig.\ref{Fig3}. At small coupling
$J_{1}/W_{1}\ll 0.33$,  $T_{c}$ is small, flat, and much extended on
the $p$ axis, qualitatively similar to the one-band
results.\cite{CHA01} However, at intermediate coupling, $T_{c}$ is
nonzero in the range from $p=0$ to $p=2x$, adopting a parabolic form
with the maximum at $p\cong x$, in contrast with the one-band model
which gives a null $T_{c}$ when $p\cong x$. The explanation is
straightforward: at $p=x$ in the one-band model the IB is fully
occupied leading to a vanishing $T_{c}$, but for the same $p$ in the
two-band model both bands are half filled, which ultimately leads to
the highest value for $T_{c}$. The $T_{c}$ dependence on the ratio
$J_{1}/W_{1}$ at some fillings $p$ is displayed in
Fig.\ref{Fig2}(b). At small coupling, $T_{c}$ correctly increases
quadratically with $J_{1}/W_{1}$, but at strong coupling $T_{c}$
incorrectly continues growing, result which will be improved upon by
the MC simulations shown below.
\begin{figure}
{\scalebox{0.32}{\includegraphics{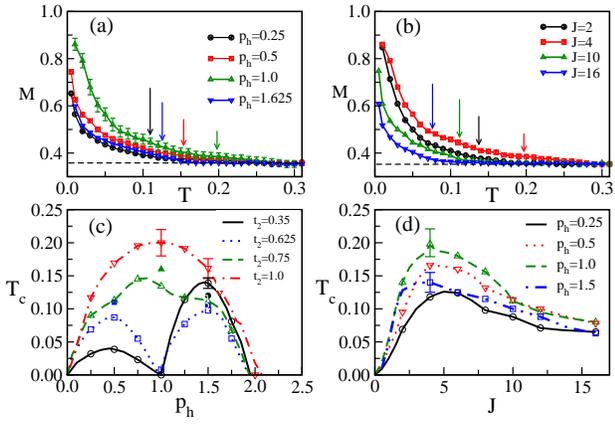}}} \caption{(Color
online) (a) MC magnetization (in absolute value) vs. temperature
($T$), with $J_1$=$J_2$=$4$ and $t_1$=$t_2$=$1$, for the hole
densities indicated. The dashed line is the exact asymptotic
high-temperature value $M_{\infty}$, which tends to 0 only in the
bulk limit. In this paper, the $T_c$ on the 5$^3$ cluster was
(arbitrarily) defined as the $T$ where $M$ reaches $\sim$5\% of the
$1-M_{\infty}$ value (indicated by arrows in (a) and (b)). Other
criteria lead to very similar conclusions. (b) Magnetization vs. $T$
for $p_h$=$1$, $t_1$=$t_2$, and several (equal) magnetic couplings
$J$; (c) Curie temperature vs. hole density for $J$=$4$ and
different ratios of the band hoppings; (d) Curie temperature versus
the (equal) magnetic interactions $J$ for several hole densities and
equal band hoppings $t_1$ and $t_2$. Results for $5^3$ ($6^3$)
lattices are indicated by open (filled) symbols. Error bars due to
the disorder (up to 7 samples) are only shown for a few points for
clarity. } \label{Fig4}
\end{figure}
%---Monte Carlo simulations ---
\section{Monte Carlo Results}
The Hamiltonian Eq.\ (\ref{ham}) was also studied numerically using
MC techniques similar to those applied to Mn-oxide
investigations.\cite{ALV03,DAG01} The fermionic sector is treated
exactly, while a MC simulation is applied to the classical localized
spins. Cubic lattices with $5^3$ and $6^3$ sites were investigated.
These lattice sizes have been shown to be sufficient for the
comparison with DMFT results and to unveil general trends. In
addition, the figures show only small variations for the $T_{c}$
estimations using the two lattices. However, if more sophisticated
quantitative analysis is required, clearly larger systems will be
needed. One may suspect that actually the number of Mn spins may
regulate the size effects, rather than the number of sites. For the
small values of $x$ used in our study, the number of Mn spins is
also very small and serious size effects could be expected. However,
in practice this does not seem to occur, and moreover the comparison
with DMFT shows similar results using both techniques. Perhaps in
the small J's regime, the delocalized nature of the carriers smears
the effects caused by the actual location of the Mn spins. These
issues deserve further study, but for our purposes of unveiling
trends in the multi-parameter space of DMS materials, the lattices
here used are sufficient.

Returning to the numerical data, the spin magnetization is the order
parameter that was used to detect the ferromagnetic
transition.\cite{DAG01} All quantities are in units of $t_1$=1, and
the density of magnetic impurities is $x\approx 0.065$. In
Fig.~4(a), typical magnetization curves are presented at several
carrier densities $p_h$, and for $J_1$=$J_2$=$J$=4 and $t_2$=$1$. In
excellent agreement with DMFT, it was observed that the estimated
$T_c$ is the highest for $p_h$=$1$. The value of $J$ used maximizes
the critical temperature, and it was confirmed that it corresponds
to the case where the IB are about to become separated from the
valence band. In Fig.~4(b), it is shown how $T_c$ increases with $J$
up to $J$=$4$, in agreement with DMFT. The strong coupling behavior
is nevertheless different since $T_c$ {\it decreases} at large $J$.
This is caused by hole localization in strong coupling, \cite{ALV03}
beyond the capabilities of DMFT. The dependence of $T_c$ on the
ratio of band hoppings is in Fig.~4(c), varying $p_h$. These results
are again in good qualitative agreement with DMFT (Fig.~2(c)). The
maximum $T_c$ for all values of $p_h$ occurs when $t_2/t_1$=$1$.
However, when $t_2$ is very different from $t_1$ the development of
magnetism is regulated by only one of the IB and the results are
similar to those obtained with a single band model, as shown in the
curves for $t_2$=$0.35$ and $0.625$ in Fig.~4(c). For $t_2/t_1$
closer to 1, a partial overlap of the IB occurs and a hump in $T_c$
develops at $p_h$=$1$ (see curve for $t_2$=$0.75$).
\begin{figure}
{\scalebox{0.32}{\includegraphics{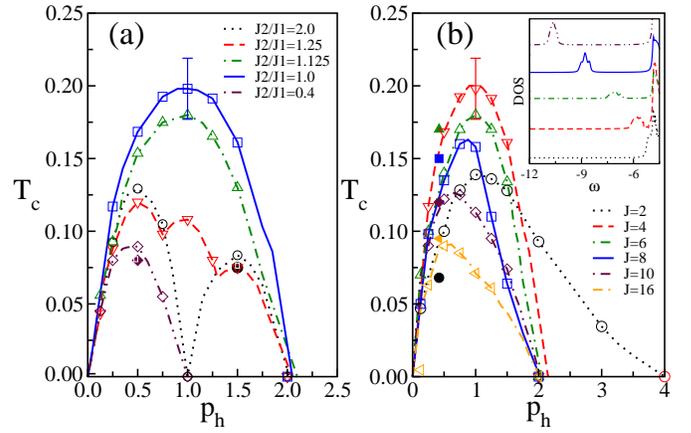}}} \caption{(Color
online) (a) $T_c$ vs. $p_h$ obtained with MC on a $5^3$ ($6^3$)
lattice with $t_1$=$t_2$=$1$ for the values of $J_2/J_1$ indicated
by the open (filled) symbols.
$J_1$ is fixed to 4, i.e. when the IB are about to separate from the
valence band for band 1 (inset Fig.~5(b)). For $J_2<J_1$ (e.g.
$J_2/J_1$=$0.4$ curve), $T_c$ is regulated by the IB in band 1,
since the IB for band 2 is deep into the valence band. In this case,
a single-band behavior is observed, with $T_c$ maximized for
$p_h$=$0.5$. For $J_2>J_1$, both IB play a role. For $J_2\gg J_1$
(see $J_2/J_1$=$2$), the two IB do not overlap, and for $0\le p_h\le
1$ $T_c$ is determined by the band-2 IB reaching a maximum at
$p_h$=$0.5$ and vanishing at $p_h$=$1$. For larger $p_h$, $T_c$ is
controlled now by the IB 1, and it raises again passing through a
maximum at $p_h$=$1.5$ and vanishing at $p_h$=$2$. For
$J_2/J_1$=$1.25$, the two IB overlap and we observe residual local
maxima at $p_h$=$0.5$ and $1.5$, related to the single band physics,
and a new local maximum at $p_h$=$1$ due to IB overlap for the
corresponding value of the chemical potential. $T_c$ at $p_h$=$0.5$
is boosted by the partial IB overlap as well. (b) $T_c$ vs. $p_h$
for $t_2/t_1$=$1$ and several $J$s. Inset: low-$T$ DOS.}
\label{Fig4dos}
\end{figure}
In Fig.~4(d) we show that, once $t_2/t_1$ is optimized, a similar
finite $J$ maximizes $T_c$ for several $p_h$'s. In all cases, the
optimal $J$ best balances the weak coupling behavior, with mobile
holes not much affected by the interaction with the Mn ions, and the
strong coupling region where the hole spins strongly align with the
Mn spins, becoming localized. This ``sweet spot'' is achieved when
the IB are about to be separated from the valence bands.

$T_c$ vs. $p_h$, at several ratios $J_2/J_1$ and for $t_1$=$t_2$ is
presented in Fig.~5(a). In  agreement with DMFT (Fig.~1), $T_c$ is
maximized for all values of $p_h$ if $J_2$=$J_1$, with the highest
value at $p_h$=$1$. Overall, there is an excellent agreement with
DMFT, as described in the caption.\cite{agreement} $T_c$ vs. $p_h$
for several $J_1$=$J_2$=$J$ is in Fig.~5(b). At small $J$, again the
MC results resemble those obtained with DMFT (Fig.~3). For, e.g.,
$J$=$2$ the IB are not formed yet (inset). In this regime, $T_c$
remains finite, although small, even for $p_h$ larger than 2.
Increasing $J$, a nonzero $T_c$ is obtained only for $p_h$ between 0
and 2, due to IB formation. $T_c$ reaches a maximum at $J$=4.

\section{Conclusions}
We have carried out the first study of a multiband model for DMS
using a powerful combination of nonperturbative techniques, DMFT and
MC. We found the parameter regime that maximizes $T_c$. This happens
at intermediate couplings and for all hole densities when
$J_{1}/J_{2}$=$1$ and $W_{1}/W_{2}$=$1$. The maximum $T_{c}$ is
obtained at $p\cong x$, in contrast with the one-band model which
has a vanishing $T_c$ at the same doping. In addition,  $T_{c}$  at
filling $p\cong x/2$ in the one-band case is smaller than with two
bands by a factor $\sim$2. In view of the simplicity of the main
results, it is clear that adding an extra band to the calculations
(which is relevant for system with negligible SO, but considerably
raises the CPU cost) will only lead to a further increase in $T_{c}$
when all the IB overlap.

The excellent agreement DMFT-MC is somewhat surprising due to the
fact that Monte Carlo considers the influence of the random location
of the Mn sites much better than DMFT. However, at small and
intermediate J's, the carriers can be sufficiently delocalized that
a smearing effect may occurs and considering the quenched disorder
only in average appears to be sufficient. Certainly at large J's the
MC and DMFT methods give totally different answers, with MC
capturing the correct localization result.

The approach described here is also quantitative. In fact, using
GaAs realistic parameters such as $p$=$0.005$ ($p_{h}$=$0.1$/Mn), a
bandwidth $\sim$ 10 eV ($t_1$$\approx$ 2.5~eV), $t_2$=(1/9)$t_1$,
and assuming $J_{1}/t_{1}$=$1$ and $J_1$=$J_2$, we obtain
$T_{c}\approx 175~K$, i.e. within the experimental range. While this
excellent agreement with experiment\cite{OHN96} may be accidental,
the trends are reliable and the result improves upon single-band
estimations. Moreover, for optimal $t_2/t_1$=1 and $J_1/t_1$=2, the
$T_c$ raises to $\sim340~K$, even at small $p_h$=0.1, setting the
upper bound for DMS under a two-band model description using a cubic
lattice.

The general qualitative picture presented here can be used to search
for DMS with even higher $T_c$'s than currently known. Our results
suggest that semiconductors with the smallest heavy to light hole
mass ratio, such as AlAs, could have the highest $T_c$ if the
couplings $J$ could be tuned to its optimal value. The present
effort paves the way toward future nonperturbative studies of DMS
models using realistic ZB lattices, and points toward procedures to
further increase the Curie temperatures.\\
\section{Acknowledgments}
We acknowledge conversations with K. Aryanpour, M. Berciu, R.S.
Fishman, J. Moreno, and J. Schliemann. This research is supported by
grant NSF-DMR-0443144. ORNL is managed by UT-Battelle, LLC under
Contract No. De-AC05-00OR22725. MC simulations used the SPF program
(http://mri-fre.ornl.gov/spf).

\suppressfloats

\begin{thebibliography}{}
\bibitem{comment} The relative position
of the bands is fixed by shifting numerically the $l$=2 valence
band such that both valence bands start at the same energy,
to mimic the degeneracy of the light
and heavy bands of GaAs at $\Gamma$.
\bibitem{agreement} This agreement is even quantitative
for the optimal $T_c$ and $J$ coupling, once the factor $\sqrt{2d}$
($d$=dimension), use to rescale the hopping in DMFT, is considered.
\end{thebibliography}
\end{document}